\documentclass[aps,prl,superscriptaddress,preprintnumbers,twocolumn,groupedaddress,showpacs,nofootinbib]{revtex4}

\usepackage{graphicx}

\def\mathswitchr#1{\relax\ifmmode{\mathrm{#1}}\else$\mathrm{#1}$\fi}
\newcommand{\PH}{\mathswitchr H}
\newcommand{\Pp}{\mathswitchr p}
\newcommand{\Pe}{\mathswitchr e}
\newcommand{\Pb}{\mathswitchr b}
\newcommand{\Pt}{\mathswitchr t}

\newcommand{\Pg}{\mathswitchr g}
\newcommand{\Pw}{\mathswitchr w}

\newcommand{\Pq}{q}
\newcommand{\Pl}{l}

\newcommand{\PW}{\mathswitchr W}
\newcommand{\PZ}{\mathswitchr Z}
\newcommand{\GF}{\mathswitch {G_\mu}}
\newcommand{\sw}{\mathswitch {s_\Pw}}

\newcommand{\kT}{\mathswitch {k_{\mathrm{T}}}}

\def\mathswitch#1{\relax\ifmmode#1\else$#1$\fi}
\newcommand{\Mt}{\mathswitch {m_\Pt}}
\newcommand{\MW}{\mathswitch {M_\PW}}
\newcommand{\MZ}{\mathswitch {M_\PZ}}
\newcommand{\MH}{\mathswitch {M_\PH}}

\newcommand{\GeV}{\unskip\,\mathrm{GeV}}

\newcommand{\fb}{\unskip\,\mathrm{fb}}

\def\reffi#1{\mbox{Fig.~\ref{#1}}}

\def\citere#1{\mbox{Ref.~\cite{#1}}}
\def\citeres#1{\mbox{Refs.~\cite{#1}}}

\newcommand{\rR}{\mathrm R}
\newcommand{\rL}{\mathrm L}
\newcommand{\rS}{\mathrm S}
\newcommand{\ri}{\mathrm i}

\newcommand{\Tevatron}{\mathrm{Tev}}
\newcommand{\LHC}{\mathrm{LHC}}
\newcommand{\LO}{\mathrm{LO}}
\newcommand{\NLO}{\mathrm{NLO}}

\def\eqintext#1{#1}

\begin{document}

\preprint{FR-PHENO-2010-041}
\preprint{PSI-PR-10-14}
\preprint{ZH-TH 19/10}
\title{\boldmath{
NLO QCD corrections to $\PW^+\PW^-\Pb\bar\Pb$ production at hadron colliders}}

\author{A.~Denner}
\affiliation{Universit\"at W\"urzburg, Institut f\"ur Theoretische Physik und Astrophysik,
97074 W\"urzburg, Germany}

\author{S.~Dittmaier}
\affiliation{Albert-Ludwigs-Universit\"at Freiburg, Physikalisches Institut, 79104 Freiburg, Germany}

\author{S.~Kallweit}
\affiliation{Paul Scherrer Institut, W\"urenlingen und Villigen, 5232 Villigen PSI, Switzerland}

\author{S.~Pozzorini}
\affiliation{Institut f\"ur Theoretische Physik, Universit\"at Z\"urich, 8057 Z\"urich, Switzerland}

\date{\today}

\begin{abstract}

Top--antitop quark pairs belong to the most abundantly produced and
precisely measurable heavy-particle signatures at hadron colliders and allow
for crucial tests of the Standard Model and new-physics searches.  Here we
report on the calculation of the next-to-leading order (NLO) QCD corrections to hadronic
\eqintext{$\PW^+\PW^-\Pb\bar\Pb$} production, which provides a complete NLO
description of the production of top--antitop pairs and their subsequent
decay into W bosons and bottom quarks, including
interferences, off-shell effects, and non-resonant backgrounds.  Numerical
predictions for the Tevatron and the LHC are presented.

\end{abstract}

\maketitle

The top quark is the heaviest of all known elementary particles and is
expected to play a key role in any theory of the flavour sector of
elementary particles.  Its precise investigation is, thus, of great
importance at the current hadron colliders Tevatron and LHC, where top
quarks are mostly produced via top--antitop \eqintext{$(\Pt\bar\Pt)$} pairs.
\looseness -1

The first step towards precise theoretical predictions for \eqintext{$\Pt\bar\Pt$}
production at hadron colliders was made already about 20~years ago with
the calculation of QCD corrections at next-to-leading-order 
(NLO)~\cite{Nason:1989zy,Beenakker:1990maa,Mangano:1991jk,Frixione:1995fj}.
Later also electroweak radiative corrections were 
calculated~\cite{Beenakker:1993yr,Moretti:2006nf,Kuhn:2006vh,Bernreuther:2006vg,Bernreuther:2008aw},
and recently important progress has been achieved both 
in the resummation of logarithmically enhanced terms~\cite{Beneke:2009rj,Czakon:2009zw,Ahrens:2010zv,Kidonakis:2010dk}
and towards the inclusion of QCD corrections at
next-to-next-to-leading-order~\cite{Dittmaier:2007wz,Kniehl:2008fd,Anastasiou:2008vd,%
Czakon:2007ej,Czakon:2007wk,Czakon:2008zk,%
Bonciani:2008az,Bonciani:2009nb,Bonciani:2010mn,
GehrmannDeRidder:2009fz,Czakon:2010td}.

The above-mentioned predictions are based on the approximation of
stable (on-shell) top quarks, i.e.\ the top-quark decays, which proceed
into pairs of W~bosons and b~quarks in the Standard Model, were ignored.
Recently also studies~\cite{Bernreuther:2004jv,Melnikov:2009dn, Bernreuther:2010ny}
at the NLO QCD level have been presented that include the top-quark
decays via a spin-correlated narrow-width approximation, i.e.\
the top quarks are still on shell.
In this letter we present first results%
\footnote{Similar results on WWbb production
have recently been shown by the HELAC-OPP collaboration at the meeting 
of the HEPTOOLS network at Granada
(see http://indico.cern.ch/conferenceDisplay.py?confId=91923).}
at NLO QCD on the
further generalization that the intermediate top quarks can be
off their mass shell, i.e.\ we consider the process of
\eqintext{$\PW^+\PW^-\Pb\bar\Pb$} production, including leptonic W-boson decays.

The reaction \eqintext{$\Pp\Pp\to\PW^+\PW^-\Pb\bar\Pb+X$} represents
one of the few remaining \eqintext{$2\to4$} LHC background processes on the Les Houches 
wishlist~\cite{Binoth:2010ra}.
While various such \eqintext{$2\to4$} NLO QCD calculations
have been performed in the recent years (see e.g.\ \citere{Binoth:2010ra} 
for a review), \eqintext{$\PW^+\PW^-\Pb\bar\Pb$} production involves the treatment of 
resonant particles for the first time in a hadron-collider environment 
on that level of complexity.
The two resonances can be consistently treated in the complex-mass
scheme that was introduced at the NLO level in the context of the 
calculation of the electroweak corrections to the processes
\eqintext{$\Pe^+\Pe^-\to\PW\PW\to4\,$fermions}~\cite{Denner:2005es,Denner:2005fg},
which was the first full NLO calculation for a \eqintext{$2\to4$} particle
process.

\begin{figure}[h]
{\includegraphics[bb=150 650 400 710, width=.48\textwidth]{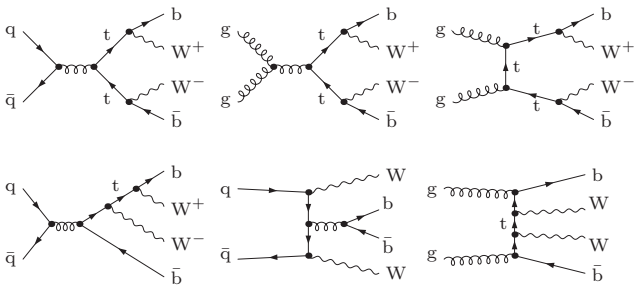}}
\vspace*{4em}
\caption{Representative LO diagrams of doubly-resonant (upper line), 
singly-resonant (first diagram in lower line), and non-resonant type (last two diagrams in lower line).}
\label{fig:treegraphs} \end{figure}

At leading order (LO), hadronic \eqintext{$\PW^+\PW^-\Pb\bar\Pb$} production proceeds
via partonic channels with quark--antiquark
(\eqintext{$\Pq\bar\Pq$}) and gluon--gluon (gg) initial states.  A few representative
diagrams are depicted in \reffi{fig:treegraphs}.  In addition to
doubly-resonant (DR) diagrams, where the \eqintext{$\PW^+\PW^-\Pb\bar\Pb$} final
state results from the decay of a  \eqintext{$\Pt\bar\Pt$} pair, our
calculation also includes singly-resonant and non-resonant contributions. 
As is well known, the bulk of the inclusive \eqintext{$\PW^+\PW^-\Pb\bar\Pb$} cross
section is efficiently reproduced by the widely used narrow-width
approximation, which incorporates all DR effects in the limit of vanishing
top-quark width, \eqintext{$\Gamma_\Pt\to 0$}.  By including all off-shell effects from
doubly-, singly-, and non-resonant diagrams, our calculation consistently
describes all contributions that are suppressed by one or more powers of
\eqintext{$\Gamma_\Pt/\Mt$}.  These extra terms are mandatory in order to achieve
percent-level precision in the (inclusive and differential) description of
\eqintext{$\Pt\bar\Pt$} production, and for a reliable simulation of off-shell
\eqintext{$\PW^+\PW^-\Pb\bar\Pb$} final states.
To describe top-quark decays in a realistic way we also include the leptonic
W-boson decays \eqintext{$\PW^+\to\nu_\Pe\Pe^+$} and \eqintext{$\PW^-\to\bar\nu_\mu\mu^-$} in
a spin-correlated narrow-width approximation.

\begin{figure} 
{\includegraphics[bb=150 650 400 710, width=.48\textwidth]{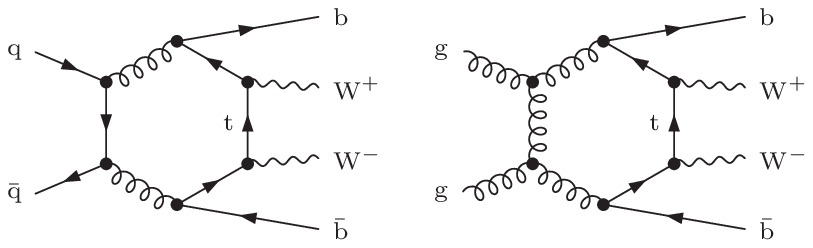}}
\vspace*{0ex}
\caption{Hexagon diagrams in \eqintext{$\Pq\bar\Pq/\Pg\Pg\to\PW^+\PW^-\Pb\bar\Pb$}.}
\label{fig:hexagons}
\end{figure}

In the following we briefly sketch the calculation of the virtual and real
corrections.  A more detailed description will be presented elsewhere.  In
order to prove the correctness of our results we have evaluated each
ingredient twice and independently.
The treatment of the virtual QCD corrections to \eqintext{$\Pq\bar\Pq/\Pg\Pg\to\PW^+\PW^-\Pb\bar{\Pb}$} 
is based on diagrammatic representations of the
one-loop amplitudes and numerical reduction of tensor
integrals~\cite{Denner:2002ii,Denner:2005nn}.  The \eqintext{$\Pq\bar\Pq$} and \eqintext{$\Pg\Pg$}
channels comprise about 300 and 800 one-loop diagrams, respectively.  The
most complicated ones are the 84 pentagons and 21 hexagons that contribute
to the gg channel (see examples in \reffi{fig:hexagons}) and
involve tensor integrals up to rank five.
Feynman diagrams are generated with two independent versions of {\sc
FeynArts}~\cite{Kublbeck:1990xc,Hahn:2000kx}, and one-loop amplitudes are
reduced along the lines of \citeres{Bredenstein:2008zb,Bredenstein:2010rs}
using two in-house {\sc Mathematica} programs, one of which relies on {\sc
FormCalc}~\cite{Hahn:1998yk} for preliminary manipulations.  The employed
approach strongly mitigates the complexity inherent in
Feynman diagrams by exploiting factorisation of colour matrices, reduction
of helicity structures to compact spinor chains, and recycling of a
multitude of common subexpressions.  
The treatment of rational
terms of ultraviolet or infrared origin is described in Appendix~A of
\citere{Bredenstein:2008zb}.
The reduced expressions are
automatically converted into {\sc Fortran77} programs that evaluate
colour/helicity summed quantities with very high CPU efficiency.
Tensor integrals are related to scalar integrals by means of numerical
algorithms that avoid instabilities from inverse Gram determinants and other
spurious singularities~\cite{Denner:2002ii,Denner:2005nn}.  

The presence of intermediate unstable top quarks in \eqintext{$\Pp\Pp\to\PW^+\PW^-\Pb\bar\Pb+X$} 
represents a non-trivial new aspect as compared to previous
NLO QCD studies of multi-particle processes.  To regularise intermediate
top-quark resonances in a gauge-invariant way we employ the complex-mass
scheme~\cite{Denner:2005fg}.  In this approach the top-quark width
$\Gamma_\Pt$ is incorporated into the definition of the renormalised
(squared) top-quark mass, \eqintext{$\mu^2_\Pt=\Mt^2-\ri\Mt\Gamma_\Pt$}.  In the
on-shell scheme this complex parameter $\mu_\Pt^2$ is identified with the
position of the pole of the top-quark propagator, and the top mass
counterterm $\delta\mu_\Pt$ is related to the top-quark self-energy at \eqintext{$p_\Pt^2=\mu^2_\Pt$}
via (see (4.25) in \citere{Denner:2005fg})
\begin{equation}\label{eq:mtct} 
\delta\mu_\Pt=\frac{\mu_\Pt}{2}\left[
\Sigma^{\Pt,\rR}(\mu^2_\Pt) +\Sigma^{\Pt,\rL}(\mu^2_\Pt)
+2\Sigma^{\Pt,\rS}(\mu^2_\Pt) \right].
\end{equation} 
We note that an expansion of the occurring self-energies around the real
point \eqintext{$p_\Pt^2=\Mt^2$} (as e.g.\ suggested in (4.27) in \citere{Denner:2005fg})
is not sufficient for NLO accuracy, because the top-quark self-energy is not
analytic at the complex pole, \eqintext{$p_\Pt^2=\mu^2_\Pt$}.
The evaluation of one-loop scalar box integrals in presence of complex
masses represents another non-trivial aspect of the complex-mass scheme.  In
our calculation we employ the results of \citere{Denner:2010tr}, where
explicit analytic continuations have been presented for all
kinematic box configurations that are relevant for physical processes.

The real corrections receive contributions from the \eqintext{$2\to 5$}
partonic processes \eqintext{$\Pg\Pg\to\PW^+\PW^-\Pb\bar\Pb\Pg$} and \eqintext{$\Pq\bar\Pq\to\PW^+\PW^-\Pb\bar\Pb \Pg$}, as well as from crossing-related \eqintext{$\Pg\Pq$} and \eqintext{$\Pg\bar\Pq$} channels.
The \eqintext{$2\to 5$} matrix elements are evaluated with {\sc Madgraph}~\cite{Alwall:2007st}
and, alternatively, using the Weyl--van-der-Waerden formalism of \citere{Dittmaier:1998nn}.
To isolate infrared divergences and cancel them analytically 
we employ in-house implementations of
the dipole subtraction formalism~\cite{Catani:1996vz}.
Specifically this is done in dimensional regularization with strictly massless light
quarks (including b quarks) and alternatively in a hybrid scheme with small quark masses
with the respective dipole subtraction terms from \citere{Catani:2002hc}

Colour and helicity correlations that enter the subtraction procedure
are generated by means of {\sc AutoDipole}~\cite{Hasegawa:2009tx} and, alternatively, in analytic form. 
To achieve sufficient numerical stability we perform the 11-dimensional phase-space
integration using multi-channel Monte Carlo techniques
with weight optimisation~\cite{Kleiss:1994qy}.
The integration of the dipole-subtracted \eqintext{$2\to 5$} contributions
is optimised by means of additional channels corresponding to the 
dipole kinematics.

In the following we present predictions for the Tevatron 
(\eqintext{$\Pp\bar\Pp$} collisions at 
1.96 TeV) and the LHC (\eqintext{$\Pp\Pp$} collisions at 7 TeV).
In NLO\,(LO) QCD we employ MSTW2008NLO\,(LO)
parton distributions~\cite{Martin:2009iq} and describe the running of the strong coupling constant $\alpha_\rS$
with two-loop\,(one-loop) accuracy, including five active flavours.
Contributions induced by the strongly suppressed bottom-quark density are neglected.
For the gauge-boson and top-quark masses we use \eqintext{$\Mt=172\GeV$},
\eqintext{$\MW=80.399\GeV$}, and \eqintext{$\MZ=91.1876\GeV$}.  The masses of all
other quarks, including b quarks, are neglected.  In view of the
negligibly small Higgs-mass dependence we adopt the \eqintext{$\MH\to\infty$} limit,
i.e.~we omit diagrams involving Higgs bosons.  The electroweak couplings
are derived from the Fermi constant
\eqintext{$\GF=1.16637\times10^{-5}\GeV^{-2}$} in the $G_\mu$-scheme, where the
sine of the mixing angle and the electromagnetic coupling read
\eqintext{$\sw^2=1-\MW^2/\MZ^2$} and
\eqintext{$\alpha=\sqrt{2}\GF\/\MW^2\sw^2/\pi$}.
For consistency, we perform the LO and NLO calculations using the top-quark
widths \eqintext{$\Gamma_{\Pt,\rm{LO}}=1.4655\GeV$} and
\eqintext{$\Gamma_{\Pt,\rm{NLO}}=1.3376\GeV$}~\cite{Jezabek:1988iv}, respectively.
Since the leptonic W-boson decay does not receive NLO QCD corrections
we employ the NLO W-boson width
\eqintext{$\Gamma_\PW=2.0997\GeV$} everywhere.
Final-state quarks and gluons with pseudo-rapidity \eqintext{$|\eta|<5$} 
are converted into infrared-safe jets using the anti-$\kT$ algorithm~\cite{Cacciari:2008gp}.
For the Tevatron\,(LHC) we set the jet-algorithm parameter \eqintext{$R=0.4\,(0.5)$} and apply the
transverse-momentum and pseudo-rapidity cuts
\eqintext{$p_{\mathrm{T,\Pb\text{-}jet}}>20\,(30)\GeV$}, \eqintext{$|\eta_{\Pb\text{-jet}}|<2.5$}. 
Moreover, we require a missing transverse momentum of
\eqintext{$p_{\mathrm{T,miss}}>25\,(20)\GeV$} and charged leptons with
\eqintext{$p_{\mathrm{T},\Pl}>20\GeV$} and \eqintext{$|\eta_{\Pl}|<2.5$}.
\looseness-1

\begin{figure}
\vspace*{0.5em}
{\includegraphics[width=.40\textwidth]{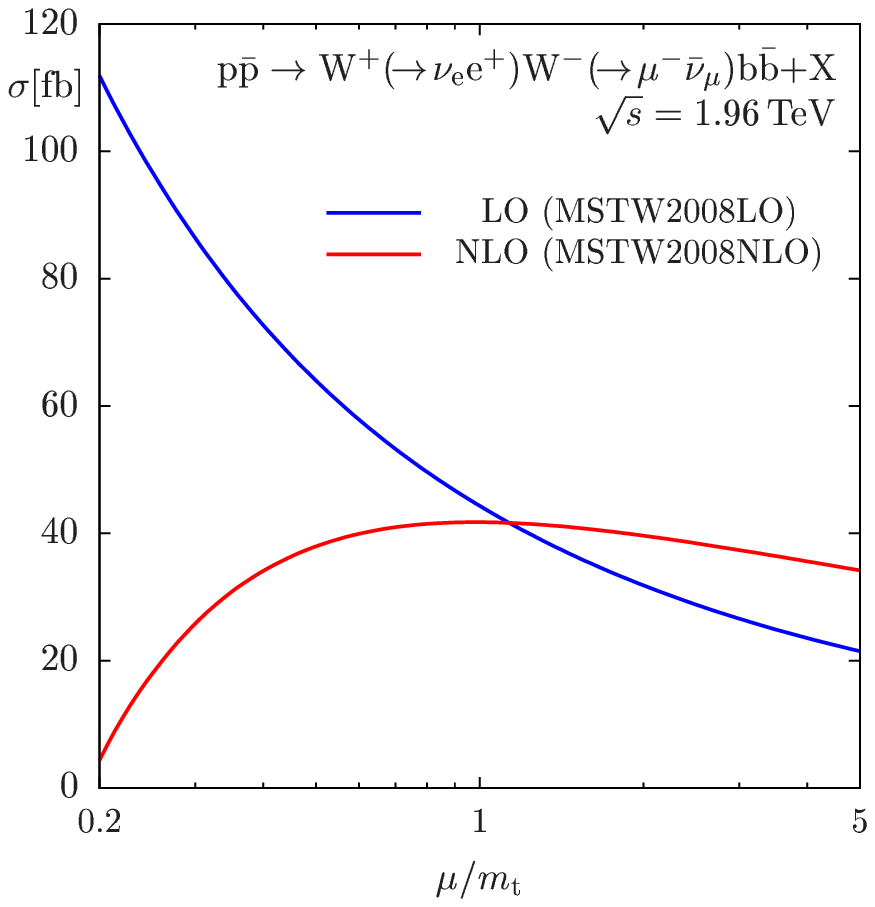}}
{\includegraphics[width=.40\textwidth]{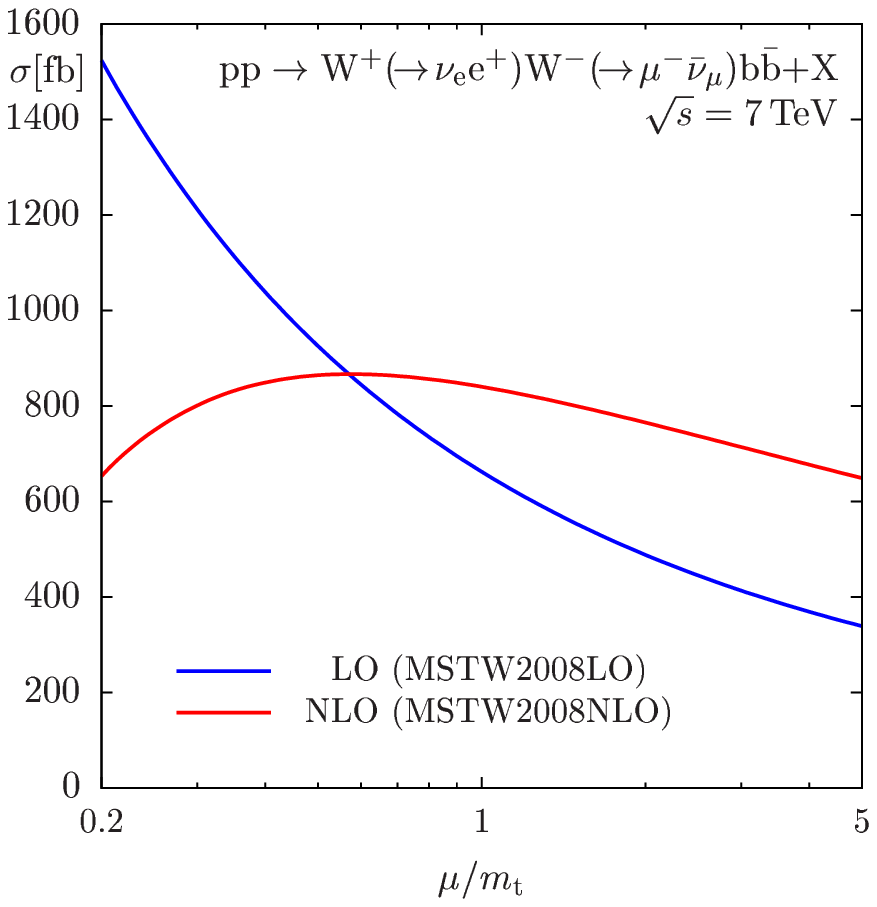}}
\vspace*{-1em}
\caption{Scale dependence of the LO and NLO \eqintext{$\PW^+\PW^-\Pb\bar\Pb$} 
 cross sections at the Tevatron and the LHC.}
\label{fig:cs-mu}
\end{figure}

The LO and NLO \eqintext{$\PW^+\PW^-\Pb\bar\Pb$} cross sections at the 
Tevatron and at the LHC are plotted in \reffi{fig:cs-mu} as a function 
of the renormalisation and factorisation scales,
\eqintext{$\mu_{\mathrm{ren}}=\mu_{\mathrm{fact}}=\mu$}.
At the Tevatron, where the cross section is dominated by the \eqintext{$\Pq\bar\Pq$} channel,
at \eqintext{$\mu=\Mt$} we obtain
\eqintext{$\sigma_{\LO}^{\Tevatron}=44.31^{+19.68}_{-12.49}\,\fb$} and
\eqintext{$\sigma_{\NLO}^{\Tevatron}=41.75^{+0.00}_{-3.79}\,\fb$}, where the uncertainties
describe missing higher-order corrections estimated via scale variations in
the range \eqintext{$\Mt/2<\mu< 2\Mt$}.  For the LHC, where the gg channel dominates, 
we obtain
\eqintext{$\sigma_{\LO}^{\LHC}=662.4^{+263.4}_{-174.1}\,\fb$} and
\eqintext{$\sigma_{\NLO}^{\LHC}=840^{+27}_{-75}\,\fb$}.  Normalising the results
to LO predictions at \eqintext{$\mu=\Mt$} we obtain the relative NLO corrections
\eqintext{$K^{\Tevatron}=0.942^{+0.000}_{-0.085}$} and
\eqintext{$K^{\LHC}=1.27^{+0.04}_{-0.11}$}.  The NLO corrections induce a moderate shift of the
integrated cross section and reduce its scale uncertainty from about 44\%\,(40\%) to 
9\%\,(9\%) at the Tevatron\,(LHC).  This confirms the good convergence of
perturbative predictions at the scale \eqintext{$\mu=\Mt$}, a feature that is reflected
also in the stable shape of the NLO curves in \reffi{fig:cs-mu}.

To assess the impact of finite-width effects on the integrated cross section
we have extrapolated our numerical results to the narrow-width limit
\eqintext{$\Gamma_\Pt\to 0$}.  In this region we observe a linear
$\Gamma_\Pt$-dependence, consistent with the cancellation of logarithmic
soft-gluon singularities.  At the Tevatron we find that finite-width effects
shift the LO(NLO) cross section by about $-0.8\%(-0.9\%)$.  At the LHC we
observe a qualitatively different behaviour: the shift induced by
finite-width contributions is smaller in size and positive.  At LO it
amounts to $+0.4\%$, and at NLO it becomes as small as the
Monte Carlo statistical error ($+0.2\%$).

\begin{figure}
\vspace*{-2.5em}
{\includegraphics[bb=150 160 450 710, width=.48\textwidth]{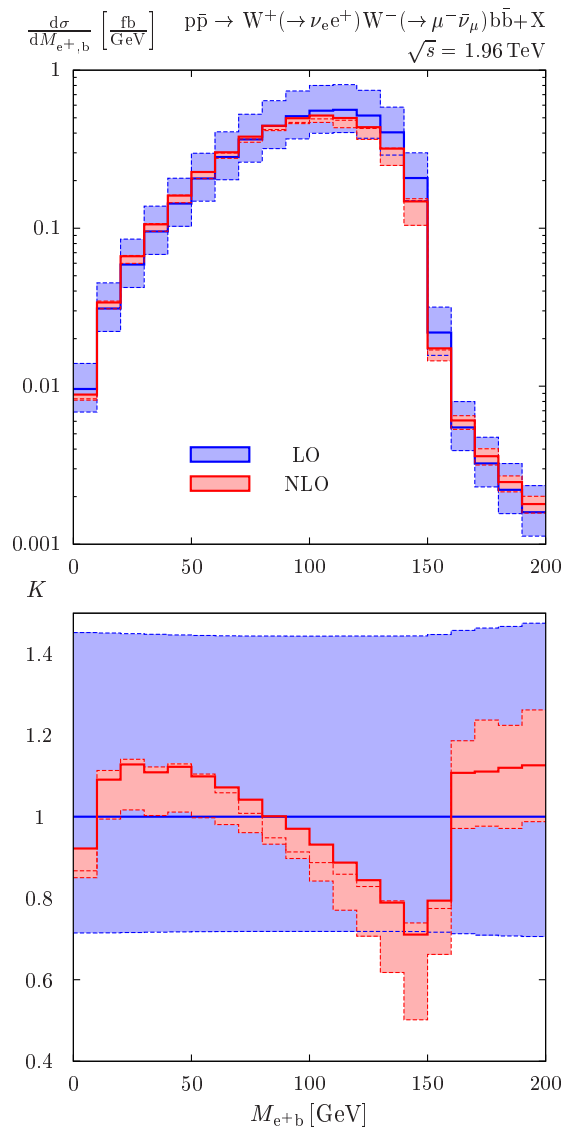}}
\vspace*{-2.0em}
\caption{Invariant mass $M_{\Pe^+\Pb}$ of the positron--b-jet system at the Tevatron:
absolute LO and NLO predictions (upper plot) and relative corrections w.r.t.~LO at
\eqintext{$\mu=\Mt$} (lower plot). The uncertainty bands describe \eqintext{$\Mt/2<\mu< 2\Mt$} variations.}
\label{fig:cs-mlb}
\end{figure}

To illustrate NLO and finite-width corrections to differential observables,
in \reffi{fig:cs-mlb} we plot the invariant-mass distribution of a positron
and a b-jet---the visible products of a top-quark decay---at the Tevatron.  In
narrow-width and LO approximation this kinematic quantity is characterised
by a sharp upper bound, \eqintext{$M^2_{\Pe^+\Pb}\le\Mt^2-\MW^2$}, which renders it
very sensitive to the top-quark mass.  The value of $\Mt$ can be
extracted with high precision using, for instance,  the invariant-mass distribution of a
positron and a $J/\psi$ from a $B$-meson decay~\cite{Kharchilava:1999yj}, an
observable that is closely related to $M_{\Pe^+\Pb}$.  In \reffi{fig:cs-mlb} we clearly see, already in LO, small but non-negligible 
off-shell contributions that elude the kinematic bound.  At NLO this feature becomes more pronounced, 
also due to QCD radiation that enters the b-jet without being emitted from its
parent b quark.  Below the kinematic bound we find very
significant NLO effects.  In the region \eqintext{$50\GeV < M_{\Pe^+\Pb}<150\GeV$}  the
shape of $M_{\Pe^+\Pb}$ is strongly distorted, with corrections
ranging from $+15\%$ to $-30\%$ (see lower plot).  In the vicinity of the kinematic bound the NLO prediction is
barely consistent with the LO uncertainty band.  This example demonstrates
the importance of \eqintext{$2\to 4$} NLO predictions for a precise description of the
kinematic details of the \eqintext{$\PW^+\PW^-\Pb\bar\Pb$} final state and, more generally,
for the top-physics programme at the Tevatron and at the LHC.
\looseness-1

\begin{acknowledgments}
S.P.\ thanks the Swiss National Science Foundation for support.
This work is supported in part by the
European Community's Marie-Curie Research Training Network 
HEPTOOLS under contract MRTN-CT-2006-035505.

Note added: Shortly after the submission of this paper results
 of a similar calculation by the HELAC-OPP Collaboration have been
published in \citere{Bevilacqua:2010qb}.

\end{acknowledgments}

\bibliography{ppwwbb_let}

\begin{thebibliography}{49}
\expandafter\ifx\csname natexlab\endcsname\relax\def\natexlab#1{#1}\fi
\expandafter\ifx\csname bibnamefont\endcsname\relax
  \def\bibnamefont#1{#1}\fi
\expandafter\ifx\csname bibfnamefont\endcsname\relax
  \def\bibfnamefont#1{#1}\fi
\expandafter\ifx\csname citenamefont\endcsname\relax
  \def\citenamefont#1{#1}\fi
\expandafter\ifx\csname url\endcsname\relax
  \def\url#1{\texttt{#1}}\fi
\expandafter\ifx\csname urlprefix\endcsname\relax\def\urlprefix{URL }\fi
\providecommand{\bibinfo}[2]{#2}
\providecommand{\eprint}[2][]{\url{#2}}

\bibitem[{\citenamefont{Nason et~al.}(1989)}]{Nason:1989zy}
\bibinfo{author}{\bibfnamefont{P.}~\bibnamefont{Nason}} \bibnamefont{et~al.},
  \bibinfo{journal}{Nucl. Phys.} \textbf{\bibinfo{volume}{B327}},
  \bibinfo{pages}{49} (\bibinfo{year}{1989}).

\bibitem[{\citenamefont{Beenakker et~al.}(1991)}]{Beenakker:1990maa}
\bibinfo{author}{\bibfnamefont{W.}~\bibnamefont{Beenakker}}
  \bibnamefont{et~al.}, \bibinfo{journal}{Nucl. Phys.}
  \textbf{\bibinfo{volume}{B351}}, \bibinfo{pages}{507} (\bibinfo{year}{1991}).

\bibitem[{\citenamefont{Mangano et~al.}(1992)\citenamefont{Mangano, Nason, and
  Ridolfi}}]{Mangano:1991jk}
\bibinfo{author}{\bibfnamefont{M.~L.} \bibnamefont{Mangano}},
  \bibinfo{author}{\bibfnamefont{P.}~\bibnamefont{Nason}}, \bibnamefont{and}
  \bibinfo{author}{\bibfnamefont{G.}~\bibnamefont{Ridolfi}},
  \bibinfo{journal}{Nucl. Phys.} \textbf{\bibinfo{volume}{B373}},
  \bibinfo{pages}{295} (\bibinfo{year}{1992}).

\bibitem[{\citenamefont{Frixione et~al.}(1995)}]{Frixione:1995fj}
\bibinfo{author}{\bibfnamefont{S.}~\bibnamefont{Frixione}}
  \bibnamefont{et~al.}, \bibinfo{journal}{Phys. Lett.}
  \textbf{\bibinfo{volume}{B351}}, \bibinfo{pages}{555} (\bibinfo{year}{1995}).

\bibitem[{\citenamefont{Beenakker et~al.}(1994)}]{Beenakker:1993yr}
\bibinfo{author}{\bibfnamefont{W.}~\bibnamefont{Beenakker}}
  \bibnamefont{et~al.}, \bibinfo{journal}{Nucl. Phys.}
  \textbf{\bibinfo{volume}{B411}}, \bibinfo{pages}{343} (\bibinfo{year}{1994}).

\bibitem[{\citenamefont{Moretti et~al.}(2006)\citenamefont{Moretti, Nolten, and
  Ross}}]{Moretti:2006nf}
\bibinfo{author}{\bibfnamefont{S.}~\bibnamefont{Moretti}},
  \bibinfo{author}{\bibfnamefont{M.~R.} \bibnamefont{Nolten}},
  \bibnamefont{and} \bibinfo{author}{\bibfnamefont{D.~A.} \bibnamefont{Ross}},
  \bibinfo{journal}{Phys. Lett.} \textbf{\bibinfo{volume}{B639}},
  \bibinfo{pages}{513} (\bibinfo{year}{2006}).

\bibitem[{\citenamefont{{K\"uhn} et~al.}(2007)\citenamefont{{K\"uhn}, Scharf,
  and Uwer}}]{Kuhn:2006vh}
\bibinfo{author}{\bibfnamefont{J.~H.} \bibnamefont{{K\"uhn}}},
  \bibinfo{author}{\bibfnamefont{A.}~\bibnamefont{Scharf}}, \bibnamefont{and}
  \bibinfo{author}{\bibfnamefont{P.}~\bibnamefont{Uwer}},
  \bibinfo{journal}{Eur. Phys. J.} \textbf{\bibinfo{volume}{C51}},
  \bibinfo{pages}{37} (\bibinfo{year}{2007}).

\bibitem[{\citenamefont{Bernreuther et~al.}(2006)\citenamefont{Bernreuther,
  {F\"ucker}, and Si}}]{Bernreuther:2006vg}
\bibinfo{author}{\bibfnamefont{W.}~\bibnamefont{Bernreuther}},
  \bibinfo{author}{\bibfnamefont{M.}~\bibnamefont{{F\"ucker}}},
  \bibnamefont{and} \bibinfo{author}{\bibfnamefont{Z.-G.} \bibnamefont{Si}},
  \bibinfo{journal}{Phys.Rev.} \textbf{\bibinfo{volume}{D74}},
  \bibinfo{pages}{113005} (\bibinfo{year}{2006}), \eprint{hep-ph/0610334}.

\bibitem[{\citenamefont{Bernreuther et~al.}(2008)\citenamefont{Bernreuther,
  {F\"ucker}, and Si}}]{Bernreuther:2008aw}
\bibinfo{author}{\bibfnamefont{W.}~\bibnamefont{Bernreuther}},
  \bibinfo{author}{\bibfnamefont{M.}~\bibnamefont{{F\"ucker}}},
  \bibnamefont{and} \bibinfo{author}{\bibfnamefont{Z.-G.} \bibnamefont{Si}},
  \bibinfo{journal}{Nuovo Cim.} \textbf{\bibinfo{volume}{B123}},
  \bibinfo{pages}{1036} (\bibinfo{year}{2008}).

\bibitem[{\citenamefont{Beneke et~al.}(2010)\citenamefont{Beneke, Falgari, and
  Schwinn}}]{Beneke:2009rj}
\bibinfo{author}{\bibfnamefont{M.}~\bibnamefont{Beneke}},
  \bibinfo{author}{\bibfnamefont{P.}~\bibnamefont{Falgari}}, \bibnamefont{and}
  \bibinfo{author}{\bibfnamefont{C.}~\bibnamefont{Schwinn}},
  \bibinfo{journal}{Nucl. Phys.} \textbf{\bibinfo{volume}{B828}},
  \bibinfo{pages}{69} (\bibinfo{year}{2010}), \eprint{0907.1443}.

\bibitem[{\citenamefont{Czakon et~al.}(2009)\citenamefont{Czakon, Mitov, and
  Sterman}}]{Czakon:2009zw}
\bibinfo{author}{\bibfnamefont{M.}~\bibnamefont{Czakon}},
  \bibinfo{author}{\bibfnamefont{A.}~\bibnamefont{Mitov}}, \bibnamefont{and}
  \bibinfo{author}{\bibfnamefont{G.~F.} \bibnamefont{Sterman}},
  \bibinfo{journal}{Phys. Rev.} \textbf{\bibinfo{volume}{D80}},
  \bibinfo{pages}{074017} (\bibinfo{year}{2009}), \eprint{0907.1790}.

\bibitem[{\citenamefont{Ahrens et~al.}(2010)}]{Ahrens:2010zv}
\bibinfo{author}{\bibfnamefont{V.}~\bibnamefont{Ahrens}} \bibnamefont{et~al.},
  \bibinfo{journal}{JHEP} \textbf{\bibinfo{volume}{1009}}, \bibinfo{pages}{097}
  (\bibinfo{year}{2010}), \eprint{1003.5827}.

\bibitem[{\citenamefont{Kidonakis}(2010)}]{Kidonakis:2010dk}
\bibinfo{author}{\bibfnamefont{N.}~\bibnamefont{Kidonakis}},
  \bibinfo{journal}{Phys.Rev.} \textbf{\bibinfo{volume}{D82}},
  \bibinfo{pages}{114030} (\bibinfo{year}{2010}), \eprint{1009.4935}.

\bibitem[{\citenamefont{Dittmaier et~al.}(2007)\citenamefont{Dittmaier, Uwer,
  and Weinzierl}}]{Dittmaier:2007wz}
\bibinfo{author}{\bibfnamefont{S.}~\bibnamefont{Dittmaier}},
  \bibinfo{author}{\bibfnamefont{P.}~\bibnamefont{Uwer}}, \bibnamefont{and}
  \bibinfo{author}{\bibfnamefont{S.}~\bibnamefont{Weinzierl}},
  \bibinfo{journal}{Phys. Rev. Lett.} \textbf{\bibinfo{volume}{98}},
  \bibinfo{pages}{262002} (\bibinfo{year}{2007}), \eprint{hep-ph/0703120}.

\bibitem[{\citenamefont{Kniehl et~al.}(2008)}]{Kniehl:2008fd}
\bibinfo{author}{\bibfnamefont{B.}~\bibnamefont{Kniehl}} \bibnamefont{et~al.},
  \bibinfo{journal}{Phys. Rev.} \textbf{\bibinfo{volume}{D78}},
  \bibinfo{pages}{094013} (\bibinfo{year}{2008}).

\bibitem[{\citenamefont{Anastasiou and Aybat}(2008)}]{Anastasiou:2008vd}
\bibinfo{author}{\bibfnamefont{C.}~\bibnamefont{Anastasiou}} \bibnamefont{and}
  \bibinfo{author}{\bibfnamefont{S.~M.} \bibnamefont{Aybat}},
  \bibinfo{journal}{Phys. Rev.} \textbf{\bibinfo{volume}{D78}},
  \bibinfo{pages}{114006} (\bibinfo{year}{2008}).

\bibitem[{\citenamefont{Czakon et~al.}(2007)\citenamefont{Czakon, Mitov, and
  Moch}}]{Czakon:2007ej}
\bibinfo{author}{\bibfnamefont{M.}~\bibnamefont{Czakon}},
  \bibinfo{author}{\bibfnamefont{A.}~\bibnamefont{Mitov}}, \bibnamefont{and}
  \bibinfo{author}{\bibfnamefont{S.}~\bibnamefont{Moch}},
  \bibinfo{journal}{Phys. Lett.} \textbf{\bibinfo{volume}{B651}},
  \bibinfo{pages}{147} (\bibinfo{year}{2007}).

\bibitem[{\citenamefont{Czakon et~al.}(2008)\citenamefont{Czakon, Mitov, and
  Moch}}]{Czakon:2007wk}
\bibinfo{author}{\bibfnamefont{M.}~\bibnamefont{Czakon}},
  \bibinfo{author}{\bibfnamefont{A.}~\bibnamefont{Mitov}}, \bibnamefont{and}
  \bibinfo{author}{\bibfnamefont{S.}~\bibnamefont{Moch}},
  \bibinfo{journal}{Nucl. Phys.} \textbf{\bibinfo{volume}{B798}},
  \bibinfo{pages}{210} (\bibinfo{year}{2008}).

\bibitem[{\citenamefont{Czakon}(2008)}]{Czakon:2008zk}
\bibinfo{author}{\bibfnamefont{M.}~\bibnamefont{Czakon}},
  \bibinfo{journal}{Phys. Lett.} \textbf{\bibinfo{volume}{B664}},
  \bibinfo{pages}{307} (\bibinfo{year}{2008}).

\bibitem[{\citenamefont{Bonciani et~al.}(2008)}]{Bonciani:2008az}
\bibinfo{author}{\bibfnamefont{R.}~\bibnamefont{Bonciani}}
  \bibnamefont{et~al.}, \bibinfo{journal}{JHEP} \textbf{\bibinfo{volume}{07}},
  \bibinfo{pages}{129} (\bibinfo{year}{2008}).

\bibitem[{\citenamefont{Bonciani et~al.}(2009)}]{Bonciani:2009nb}
\bibinfo{author}{\bibfnamefont{R.}~\bibnamefont{Bonciani}}
  \bibnamefont{et~al.}, \bibinfo{journal}{JHEP}
  \textbf{\bibinfo{volume}{0908}}, \bibinfo{pages}{067} (\bibinfo{year}{2009}),
  \eprint{0906.3671}.

\bibitem[{\citenamefont{Bonciani et~al.}(2010)}]{Bonciani:2010mn}
\bibinfo{author}{\bibfnamefont{R.}~\bibnamefont{Bonciani}} \bibnamefont{et~al.}
  (\bibinfo{year}{2010}), \eprint{1011.6661}.

\bibitem[{\citenamefont{Gehrmann-De~Ridder and
  Ritzmann}(2009)}]{GehrmannDeRidder:2009fz}
\bibinfo{author}{\bibfnamefont{A.}~\bibnamefont{Gehrmann-De~Ridder}}
  \bibnamefont{and} \bibinfo{author}{\bibfnamefont{M.}~\bibnamefont{Ritzmann}},
  \bibinfo{journal}{JHEP} \textbf{\bibinfo{volume}{0907}}, \bibinfo{pages}{041}
  (\bibinfo{year}{2009}), \eprint{0904.3297}.

\bibitem[{\citenamefont{Czakon}(2010)}]{Czakon:2010td}
\bibinfo{author}{\bibfnamefont{M.}~\bibnamefont{Czakon}},
  \bibinfo{journal}{Phys. Lett.} \textbf{\bibinfo{volume}{B693}},
  \bibinfo{pages}{259} (\bibinfo{year}{2010}), \eprint{1005.0274}.

\bibitem[{\citenamefont{Bernreuther et~al.}(2004)}]{Bernreuther:2004jv}
\bibinfo{author}{\bibfnamefont{W.}~\bibnamefont{Bernreuther}}
  \bibnamefont{et~al.}, \bibinfo{journal}{Nucl. Phys.}
  \textbf{\bibinfo{volume}{B690}}, \bibinfo{pages}{81} (\bibinfo{year}{2004}),
  \eprint{hep-ph/0403035}.

\bibitem[{\citenamefont{Melnikov and Schulze}(2009)}]{Melnikov:2009dn}
\bibinfo{author}{\bibfnamefont{K.}~\bibnamefont{Melnikov}} \bibnamefont{and}
  \bibinfo{author}{\bibfnamefont{M.}~\bibnamefont{Schulze}},
  \bibinfo{journal}{JHEP} \textbf{\bibinfo{volume}{08}}, \bibinfo{pages}{049}
  (\bibinfo{year}{2009}).

\bibitem[{\citenamefont{Bernreuther and Si}(2010)}]{Bernreuther:2010ny}
\bibinfo{author}{\bibfnamefont{W.}~\bibnamefont{Bernreuther}} \bibnamefont{and}
  \bibinfo{author}{\bibfnamefont{Z.-G.} \bibnamefont{Si}},
  \bibinfo{journal}{Nucl. Phys.} \textbf{\bibinfo{volume}{B837}},
  \bibinfo{pages}{90} (\bibinfo{year}{2010}).

\bibitem[{\citenamefont{Andersen et~al.}(2010)}]{Binoth:2010ra}
\bibinfo{author}{\bibfnamefont{J.~R.} \bibnamefont{Andersen}}
  \bibnamefont{et~al.} (\bibinfo{collaboration}{SM and NLO Multileg Working
  Group}) (\bibinfo{year}{2010}), \eprint{1003.1241}.

\bibitem[{\citenamefont{Denner et~al.}(2005{\natexlab{a}})}]{Denner:2005es}
\bibinfo{author}{\bibfnamefont{A.}~\bibnamefont{Denner}} \bibnamefont{et~al.},
  \bibinfo{journal}{Phys. Lett.} \textbf{\bibinfo{volume}{B612}},
  \bibinfo{pages}{223} (\bibinfo{year}{2005}{\natexlab{a}}).

\bibitem[{\citenamefont{Denner et~al.}(2005{\natexlab{b}})}]{Denner:2005fg}
\bibinfo{author}{\bibfnamefont{A.}~\bibnamefont{Denner}} \bibnamefont{et~al.},
  \bibinfo{journal}{Nucl. Phys.} \textbf{\bibinfo{volume}{B724}},
  \bibinfo{pages}{247} (\bibinfo{year}{2005}{\natexlab{b}}).

\bibitem[{\citenamefont{Denner and Dittmaier}(2003)}]{Denner:2002ii}
\bibinfo{author}{\bibfnamefont{A.}~\bibnamefont{Denner}} \bibnamefont{and}
  \bibinfo{author}{\bibfnamefont{S.}~\bibnamefont{Dittmaier}},
  \bibinfo{journal}{Nucl. Phys.} \textbf{\bibinfo{volume}{B658}},
  \bibinfo{pages}{175} (\bibinfo{year}{2003}).

\bibitem[{\citenamefont{Denner and Dittmaier}(2006)}]{Denner:2005nn}
\bibinfo{author}{\bibfnamefont{A.}~\bibnamefont{Denner}} \bibnamefont{and}
  \bibinfo{author}{\bibfnamefont{S.}~\bibnamefont{Dittmaier}},
  \bibinfo{journal}{Nucl. Phys.} \textbf{\bibinfo{volume}{B734}},
  \bibinfo{pages}{62} (\bibinfo{year}{2006}).

\bibitem[{\citenamefont{{K\"ublbeck} et~al.}(1990)\citenamefont{{K\"ublbeck},
  {B\"ohm}, and Denner}}]{Kublbeck:1990xc}
\bibinfo{author}{\bibfnamefont{J.}~\bibnamefont{{K\"ublbeck}}},
  \bibinfo{author}{\bibfnamefont{M.}~\bibnamefont{{B\"ohm}}}, \bibnamefont{and}
  \bibinfo{author}{\bibfnamefont{A.}~\bibnamefont{Denner}},
  \bibinfo{journal}{Comput. Phys. Commun.} \textbf{\bibinfo{volume}{60}},
  \bibinfo{pages}{165} (\bibinfo{year}{1990}).

\bibitem[{\citenamefont{Hahn}(2001)}]{Hahn:2000kx}
\bibinfo{author}{\bibfnamefont{T.}~\bibnamefont{Hahn}},
  \bibinfo{journal}{Comput. Phys. Commun.} \textbf{\bibinfo{volume}{140}},
  \bibinfo{pages}{418} (\bibinfo{year}{2001}).

\bibitem[{\citenamefont{Bredenstein et~al.}(2008)}]{Bredenstein:2008zb}
\bibinfo{author}{\bibfnamefont{A.}~\bibnamefont{Bredenstein}}
  \bibnamefont{et~al.}, \bibinfo{journal}{JHEP} \textbf{\bibinfo{volume}{08}},
  \bibinfo{pages}{108} (\bibinfo{year}{2008}).

\bibitem[{\citenamefont{Bredenstein et~al.}(2010)}]{Bredenstein:2010rs}
\bibinfo{author}{\bibfnamefont{A.}~\bibnamefont{Bredenstein}}
  \bibnamefont{et~al.}, \bibinfo{journal}{JHEP}
  \textbf{\bibinfo{volume}{1003}}, \bibinfo{pages}{021} (\bibinfo{year}{2010}).

\bibitem[{\citenamefont{Hahn and Perez-Victoria}(1999)}]{Hahn:1998yk}
\bibinfo{author}{\bibfnamefont{T.}~\bibnamefont{Hahn}} \bibnamefont{and}
  \bibinfo{author}{\bibfnamefont{M.}~\bibnamefont{Perez-Victoria}},
  \bibinfo{journal}{Comput. Phys. Commun.} \textbf{\bibinfo{volume}{118}},
  \bibinfo{pages}{153} (\bibinfo{year}{1999}).

\bibitem[{\citenamefont{Denner and Dittmaier}(2011)}]{Denner:2010tr}
\bibinfo{author}{\bibfnamefont{A.}~\bibnamefont{Denner}} \bibnamefont{and}
  \bibinfo{author}{\bibfnamefont{S.}~\bibnamefont{Dittmaier}},
  \bibinfo{journal}{Nucl. Phys.} \textbf{\bibinfo{volume}{B844}},
  \bibinfo{pages}{199} (\bibinfo{year}{2011}).

\bibitem[{\citenamefont{Alwall et~al.}(2007)}]{Alwall:2007st}
\bibinfo{author}{\bibfnamefont{J.}~\bibnamefont{Alwall}} \bibnamefont{et~al.},
  \bibinfo{journal}{JHEP} \textbf{\bibinfo{volume}{09}}, \bibinfo{pages}{028}
  (\bibinfo{year}{2007}).

\bibitem[{\citenamefont{Dittmaier}(1999)}]{Dittmaier:1998nn}
\bibinfo{author}{\bibfnamefont{S.}~\bibnamefont{Dittmaier}},
  \bibinfo{journal}{Phys. Rev.} \textbf{\bibinfo{volume}{D59}},
  \bibinfo{pages}{016007} (\bibinfo{year}{1999}).

\bibitem[{\citenamefont{Catani and Seymour}(1997)}]{Catani:1996vz}
\bibinfo{author}{\bibfnamefont{S.}~\bibnamefont{Catani}} \bibnamefont{and}
  \bibinfo{author}{\bibfnamefont{M.~H.} \bibnamefont{Seymour}},
  \bibinfo{journal}{Nucl. Phys.} \textbf{\bibinfo{volume}{B485}},
  \bibinfo{pages}{291} (\bibinfo{year}{1997}).

\bibitem[{\citenamefont{Catani et~al.}(2002)}]{Catani:2002hc}
\bibinfo{author}{\bibfnamefont{S.}~\bibnamefont{Catani}} \bibnamefont{et~al.},
  \bibinfo{journal}{Nucl. Phys.} \textbf{\bibinfo{volume}{B627}},
  \bibinfo{pages}{189} (\bibinfo{year}{2002}).

\bibitem[{\citenamefont{Hasegawa et~al.}(2010)\citenamefont{Hasegawa, Moch, and
  Uwer}}]{Hasegawa:2009tx}
\bibinfo{author}{\bibfnamefont{K.}~\bibnamefont{Hasegawa}},
  \bibinfo{author}{\bibfnamefont{S.}~\bibnamefont{Moch}}, \bibnamefont{and}
  \bibinfo{author}{\bibfnamefont{P.}~\bibnamefont{Uwer}},
  \bibinfo{journal}{Comput. Phys. Commun.} \textbf{\bibinfo{volume}{181}},
  \bibinfo{pages}{1802} (\bibinfo{year}{2010}).

\bibitem[{\citenamefont{Kleiss and Pittau}(1994)}]{Kleiss:1994qy}
\bibinfo{author}{\bibfnamefont{R.}~\bibnamefont{Kleiss}} \bibnamefont{and}
  \bibinfo{author}{\bibfnamefont{R.}~\bibnamefont{Pittau}},
  \bibinfo{journal}{Comput. Phys. Commun.} \textbf{\bibinfo{volume}{83}},
  \bibinfo{pages}{141} (\bibinfo{year}{1994}).

\bibitem[{\citenamefont{Martin et~al.}(2009)}]{Martin:2009iq}
\bibinfo{author}{\bibfnamefont{A.~D.} \bibnamefont{Martin}}
  \bibnamefont{et~al.}, \bibinfo{journal}{Eur. Phys. J.}
  \textbf{\bibinfo{volume}{C63}}, \bibinfo{pages}{189} (\bibinfo{year}{2009}).

\bibitem[{\citenamefont{Jezabek and {K\"uhn}}(1989)}]{Jezabek:1988iv}
\bibinfo{author}{\bibfnamefont{M.}~\bibnamefont{Jezabek}} \bibnamefont{and}
  \bibinfo{author}{\bibfnamefont{J.~H.} \bibnamefont{{K\"uhn}}},
  \bibinfo{journal}{Nucl. Phys.} \textbf{\bibinfo{volume}{B314}},
  \bibinfo{pages}{1} (\bibinfo{year}{1989}).

\bibitem[{\citenamefont{Cacciari et~al.}(2008)\citenamefont{Cacciari, Salam,
  and Soyez}}]{Cacciari:2008gp}
\bibinfo{author}{\bibfnamefont{M.}~\bibnamefont{Cacciari}},
  \bibinfo{author}{\bibfnamefont{G.~P.} \bibnamefont{Salam}}, \bibnamefont{and}
  \bibinfo{author}{\bibfnamefont{G.}~\bibnamefont{Soyez}},
  \bibinfo{journal}{JHEP} \textbf{\bibinfo{volume}{04}}, \bibinfo{pages}{063}
  (\bibinfo{year}{2008}).

\bibitem[{\citenamefont{Kharchilava}(2000)}]{Kharchilava:1999yj}
\bibinfo{author}{\bibfnamefont{A.}~\bibnamefont{Kharchilava}},
  \bibinfo{journal}{Phys. Lett.} \textbf{\bibinfo{volume}{B476}},
  \bibinfo{pages}{73} (\bibinfo{year}{2000}).

\bibitem[{\citenamefont{Bevilacqua et~al.}(2010)}]{Bevilacqua:2010qb}
\bibinfo{author}{\bibfnamefont{G.}~\bibnamefont{Bevilacqua}}
  \bibnamefont{et~al.} (\bibinfo{year}{2010}), \eprint{1012.4230}.

\end{thebibliography}

\end{document}